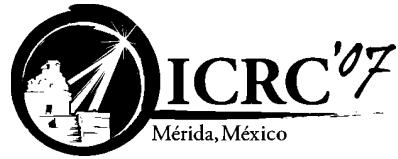

# Powerful nanosecond light sources based on LEDs for astroparticle physics experiments

B.K. LUBSANDORZHIEV[1], R.V. POLESHUK[1], B.A.J. SHAIBONOV[1], Y.E. VYATCHIN[1]

[1]*Institute for Nuclear Research of Russian Academy of Science, Moscow, Russia*
lubsand@pcbai10.inr.ruhep.ru

**Abstract:** Powerful nanosecond light sources based on LEDs have been developed for use in astroparticle physics experiments. The light sources use either matrixes of ultra bright blue InGaN LEDs or a new generation high power blue LEDs. It's shown that such light sources have light yield of up to $10^{10}$ - $10^{12}$ photons per pulse with very fast light emission kinetics. Described light sources are important for use in calibration systems of Cherenkov and scintillator detectors. The developed light sources are currently used successfully in a number of astroparticle physics experiments, namely: the TUNKA EAS experiment, the Baikal neutrino experiment etc.

## Introduction

Ultra bright blue and violet LEDs have been getting rampant development for the last decade. The advances made in this field are really stunning. Just a mere list of the LEDs manufacturers is quite impressive. The LEDs are successfully required for a number of applications. Among the main fields of applications are displays, computers car electronics etc. As for the high energy and astroparticle physics experiments the LEDs are very useful for timing and amplitude calibration systems. With special drivers it is possible to cover using just a single piece of such LED a rather big range in light pulses amplitude staying still in a few nanosecond time domain. It was shown in [1, 2] that a single ultra bright blue LED one can reach a light yield of $10^9$ photons per pulse. For this kind of application it's very important to have powerful (up to $10^8 - 10^{10}$ photons per pulse) adjustable light sources with light pulses width as short as possible (1-2 ns typically). For a large scale experiments there is a necessity to have light sources with higher light yield.

It was shown in [1, 2] also that a single ultra bright blue LEDs can withstand quite safely nanosecond current pulses with amplitude of up to 3A. The light yield of the LEDs increases almost linearly with increase of amplitude of current pulses. Unfortunately it's no possibilities to increase further current running through the LEDs for safety reasons. But how to increase light yield of light sources based on InGaN LEDs without substantial deteriorating their light emission kinetics?

## Light sources based on ultra bright blue LEDs matrixes (I).

The first idea to increase the light yield is to assemble LEDs in a matrix. There are two ways to do it. The first one is to make a matrix of LED drivers where each LED of the matrix has its own driver. The second one is to drive a matrix of LEDs with one driver.

In the first case the most difficult technical problem is to reach a high level of firing simultaneity of individual drivers and LEDs of the matrix. In order to have the fastest light emission kinetics and the largest light yield of the whole matrix one should select thoroughly individual LEDs and tune elaborately individual drivers of the matrix. Taking into account the fact that even ultra bright blue LEDs of the same type can differ very much in their light emission kinetics and light yield, to reach the high level of simultaneity and equal light yield is quite a job.

We found a number of types of ultra bright blue LEDs having a high level of repetitiveness of their characteristics. We selected the fastest LEDs without slow components and with highest light yield to use in the light sources based on matrixes of LEDs. The LEDs produced by NICHIA, KINGBRIGHT, YolDal, Bright LEDs and G-nor are among the selected types of LEDs. For each type of LEDs we selected thoroughly LEDs with very similar light emission kinetics and light yield.

As for electronic drivers for the light source it should be noted that only the drivers based on avalanche transistors can be used to reach the highest level of light yield. We use ZETEX avalanche transistors in drivers. In Fig.1 the light source based on the matrix of LED drivers is shown.

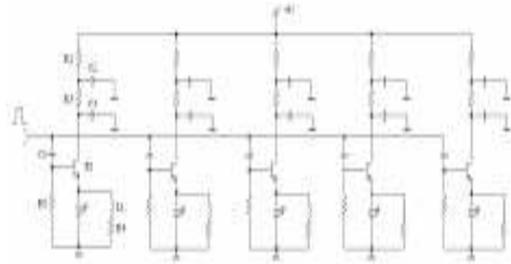

Fig.1. Scheme of a powerful nanosecond light source based on the matrix of LED drivers. The "one LED – one driver" approach.

The LEDs and drivers of the matrix are matched as precisely as possible to each other so the simultaneity of their firing is very high. The accuracy of the matrix LEDs light pulses coincidence in time is well less than 50 ps. In Fig.2 it is presented the light emission kinetics of the individual LEDs of the LED matrix shown in Fig.1. The light emission kinetics of each LED of the matrix is depicted in the figure by different colored curves. The black curve corresponds to the light emission kinetics of the whole LED matrix. One can see from the figure that the deviation of temporal behavior of each LED in the matrix is very small and resulting shape of the matrix light pulse practically follows the light pulses shape of constituting LEDs. The LEDs in the matrix are the "old" NSPB500S produced by NICHIA. As it was demonstrated in [2] the "old" NICHIA LEDs are much faster than the "new" ones.

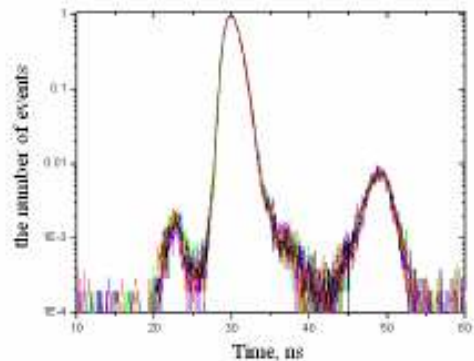

Fig.2. Light emission kinetics of the LED matrix (the black curve) and individual LEDs of the matrix (the colored curves). The "one LED – one driver" approach.

The light emission kinetics of the LEDs and the matrix was measured by a time correlated single photon counting (TCSPC) technique [3]. The light emission kinetics of each individual LED was measured by a fast PMT when all other LEDs of the matrix were masked. A set of neutral density filters was used to decrease the level of the PMT's photocathode illumination to the single photoelectron level as required by TCSPC technique. The left and right peaks in fig.2 correspond to prepulses and late pulses of the PMT used in the measurement and have no relation with the LEDs light pulses. To infer the LEDs light pulses widths one should analyze the main peak around 30 ns in fig.2, subtracting the photoelectron transit time spread (TTS) of the PMT. The TTS of the PMT was about 200-300 ps, giving negligible contribution to the width of pulses shown in fig.2.

Practically the same results are reached with the YolDal, LIGITEK and G-nor ultra bright blue LEDs. The latter are the fastest ones being a bit inferior to others by their light yield.

It should be noticed here that the above described approach to make the LEDs matrix is the most difficult one. The LEDs and electronic components of the driver (avalanche transistors, discharge capacitors, resistors and PCB boards) should be selected thoroughly. The drivers and LEDs should be tuned and matched finely to each

other – the avalanche transistors should have very close avalanche breakdown voltages, the rise time of resulting current pulses should be very close, the nominals of discharge capacitors should be nearly identical, emission kinetics and light yield of the LEDs should be identical too etc. Despite the difficulties it is feasible, in principle, to assemble such matrixes with LEDs as many as possible without substantial deteriorating light emission kinetics of the matrixes.

## Light sources based on ultra bright blue LEDs matrixes (II).

Another approach to make powerful nanosecond light sources based on matrixes of LEDs is to drive several LEDs switched in parallel in one matrix with only one driver. In this case there is necessary to select just fast LEDs with identical light emission kinetics.

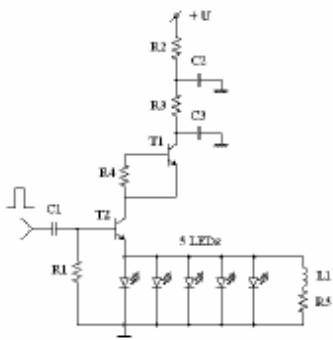

Fig.3. Scheme of a powerful nanosecond light source based on the matrix of LED drivers. The "one driver – a matrix of LEDs" approach.

The electrical scheme of such a light source is shown in Fig.3. A matrix of LEDs, several LEDs switched in parallel, is inserted into emitter circuit of the driver. A LR filter is switched in parallel to LEDs to cancel the long tail of C3 capacitor's discharge. To increase current pulse running through LEDs two avalanche transistors, the same ZETEX transistors, consequently switched are used in the driver. As a result the driver needs higher power supply voltage – $\geq 600$ V.

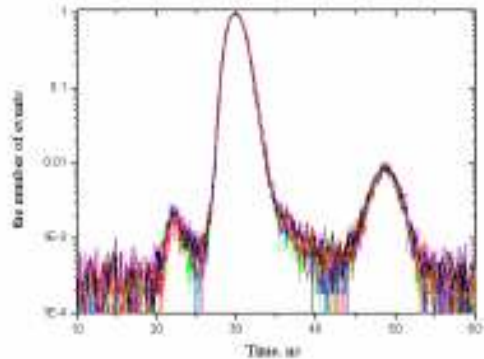

Fig.4. Light emission kinetics of the LEDs matrix (the black curve) and individual LEDs of the matrix (the colored curves). The "one driver – a matrix of LEDs" approach.

Fig.4 presents light pulses temporal profiles of the LED matrix and individual LEDs of the matrix. As in the previous case, the black curve in the figure corresponds to the whole matrix and other colored curves to the light emission kinetics of the individual LEDs of the matrix measured by the same TCSPC technique and LED masking. Once again it is seen that the LEDs of the matrix are almost ideally matched by their emission kinetics and light yield. The coincidence accuracies of the matrix LEDs light pulses in both cases are better than 50 ps.

The second approach provides 2-3 times less light yield in comparison with the first one but it is much easier to implement technically. There is no need to tune and match different drivers. One should just select properly LEDs. The light yield reached with the driver shown schematically in Fig.3 is $\sim 5\times10^9$ photons per pulse. The light emission kinetics is almost the same as in the first case. Further increase in the light yield is possible with use of a cluster of several matrixes of LEDs as demonstrated in Fig.5. With such clusters it is possible to reach the light yield of $\geq 10^{11}$ photons per pulse and at the same time staying still in the time domain of 1-2 ns light pulse width (FWHM).

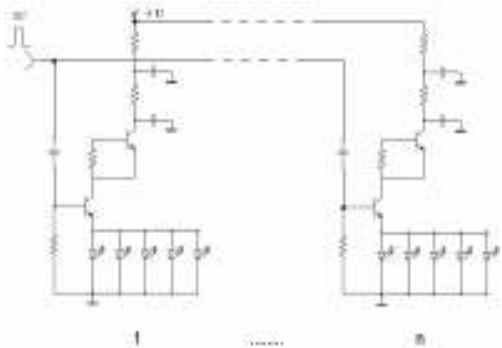

Fig.5. Cluster of *n* matrixes of LEDs.

**Light sources based on high power LEDs.**

It is quite interesting to use a relatively new player in the field – high power blue LEDs which can replace described LED matrixes in the powerful nanosecond light sources. High power blue LEDs can withstand DC current of 1A providing much higher light yield in comparison with ultra bright blue LEDs. The Fig.6 and Fig.7 presents the scheme of the driver and light pulse shape of the LXHL-NB98 high power blue LED produced by LUMILED correspondingly.

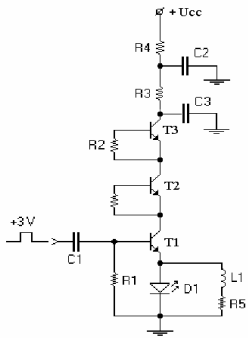

Fig.6. Scheme of a powerful nanosecond light source based on high power blue LED – LXHL-NB98.

The LEDs need highier high power supply voltage of $\geq 10^3$V. The light sources based on a single such LEDs provide the light yield of $\geq 10^{12}$ photons per pulse. We tested high power blue LEDs from LUMILED, Cree and G-nor. Their light emission kinetics are rather slow with ~5-10 ns width (FWHM) as shown in Fig.7.

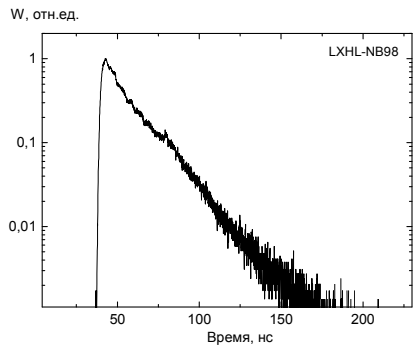

Fig.7. Light pulse shape of LXHL-NB98

**Conclusion**

Ultra bright blue LEDs give excellent opportunities to build powerful, fast and inexpensive light sources for calibration systems of astroparticle physics experiments. With matrixes of ultra bright blue LEDs it is possible to have light sources with light pulses width of 1-2 ns (FWHM) and light yield of up to $10^{10}$ photons per pulse and even more, and with a cluster of matrixes – $10^{11}$ photons per pulse without deteriorating light emission kinetics. New high power blue LEDs allow to have light sources with intensities of $10^{12}$ photons per pulse with a single such LED but their emission kinetics relatively slow with light pulses width of ~5-10 ns. Powerful nanosecond light sources based on ultra bright blue LEDs have very high long term stability. So they are in many respects very good competitor to the laser systems used in calibration systems of many astroparticle physics experiments.

**Acknowledgements**

The authors are indebted very much to Dr. V.Ch.Lubsandorzhieva for careful reading of the manusciept of the paper and many invaluable remarks.